# Leveraging the Channel as a Sensor: Real-time Vehicle Classification Using Multidimensional Radio-fingerprinting

Benjamin Sliwa[1], Nico Piatkowski[2], Marcus Haferkamp[1], Dennis Dorn[3] and Christian Wietfeld[1]

***Abstract*— Upcoming Intelligent Transportation Systems (ITSs) will transform roads from static resources to dynamic Cyber Physical Systems (CPSs) in order to satisfy the requirements of future vehicular traffic in smart city environments. Up-to-date information serves as the basis for changing street directions as well as guiding individual vehicles to a fitting parking slot. In this context, not only abstract indicators like traffic flow and density are required, but also data about mobility parameters and class information of individual vehicles. Consequently, accurate and reliable systems that are capable of providing these kinds of information in real-time are highly demanded. In this paper, we present a system for classifying vehicles based on their radio-fingerprints which applies cutting-edge machine learning models and can be non-intrusively installed into the existing road infrastructure in an ad-hoc manner. In contrast to other approaches, it is able to provide accurate classification results without causing privacy-violations or being vulnerable to challenging weather conditions. Moreover, it is a promising candidate for large-scale city deployments due to its cost-efficient installation and maintenance properties. The proposed system is evaluated in a comprehensive field evaluation campaign within an experimental live deployment on a German highway, where it is able to achieve a binary classification success ratio of more than 99% and an overall accuracy of 89.15% for a fine-grained classification task with nine different classes.**

## I. Introduction

Apart from utilizing abstract information like traffic flow and traffic density, upcoming data-driven ITSs [1] will highly benefit from integrating knowledge about the vehicle types into their decision processes for traffic control. This includes smart parking applications, where vehicles are routed to a free parking slot depending on their shape-specific spacial requirements as well as dynamic lane assignments and emission control for trucks [2]. Consequently, precise vehicle classification systems are highly demanded. Moreover, for city-based large-scale deployments, the cost-efficiency for installation and maintenance is another crucial factor. In this paper, we present a vehicle classification approach that leverages the vehicle-specific *radio-fingerprint* which is derived from the attenuation characteristics of the vehicle passing an installation of multiple radio links. In contrast to existing approaches, the proposal simultaneously fulfills strict requirements like cost-efficiency, non-intrusiveness, real-time classification capability, weather independency and privacy-preservation while achieving highly accurate classification results. Consequently, it is a promising candidate for being used in future smart city ITS. The proposed system evolved from the Wireless Detection and Warning System (WDWS) presented in [3] for the detection of wrong-way-drivers on highways and was later extended by first generic vehicle classification algorithms using a single direct radio link and ray-tracing simulations in [4]. In this paper, we move another step forward by exploiting data from all available communication links and apply cutting-edge machine-learning algorithms in order to achieve more precise and reliable classification results. Fig. 1 illustrates an example scenario where the proposed sensor system is utilized for an on-site use-case (parking space accounting) and serves as a data source for further cloud-based ITSs services. The remainder of the paper is structured as follows. After discussing relevant state-of-the-art approaches for vehicle detection and classification, we give an overview of the proposed system and provide detailed descriptions of the individual components. In the following section, we explain the experimental setup for the field measurements. Finally, detailed classification results are presented and the different data analysis methods are compared with respect to their resource-efficiency for being used on embedded devices.

## II. Related Work

In this section, we present and discuss state-of-the-art approaches for vehicle classification. Tab. I summarizes the key properties of different solution approaches and their respective vulnerabilities to environmental conditions (e.g. dirt, lightning, rainfall) and special traffic situations. Although standardized taxonomies such as the mostly axle-

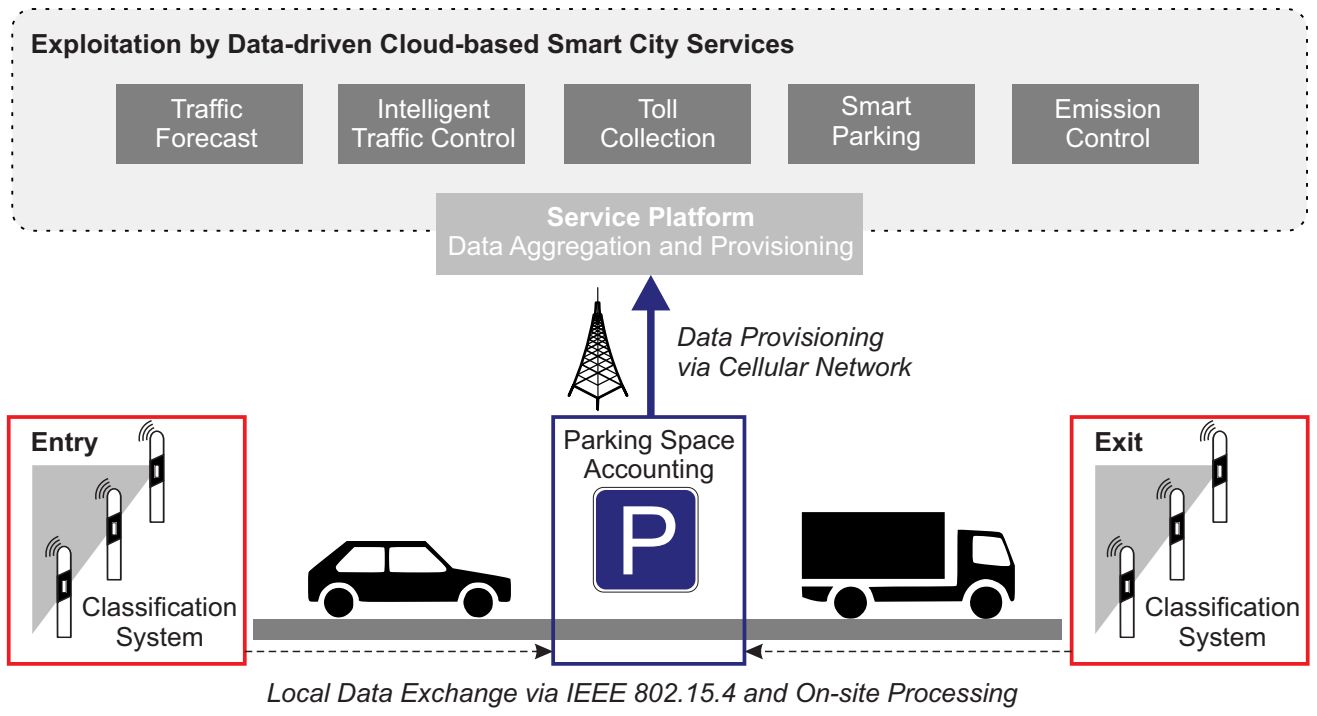

Fig. 1. Example application scenario for the proposed classification system: Two system installations detect vehicles entering and leaving a parking space. The data is utilized locally for accounting and additionally forwarded to a cloud-based service platform for data aggregation that provides the information for further exploitation by services in a smart city context.

[1]Benjamin Sliwa, Marcus Haferkamp and Christian Wietfeld are with Communication Networks Institute, TU Dortmund University, 44227 Dortmund, Germany {Benjamin.Sliwa, Marcus.Haferkamp, Christian.Wietfeld}@tu-dortmund.de

[2]Nico Piatkowski is with the Computer Science Department, AI Group, TU Dortmund University, 44227 Dortmund, Germany Nico.Piatkowski@tu-dortmund.de

[3]Dennis Dorn is with Wilhelm Schröder GmbH, 58849 Herscheid, Germany Dorn@mfds.eu

TABLE I
OVERVIEW OF STATE-OF-THE-ART APPROACHES FOR VEHICLE DETECTION AND CLASSIFICATION

| Approach | Accuracy / Classes | Cost | Intrusive | Online | Privacy | Vulnerability |
|---|---|---|---|---|---|---|
| Acoustics | 73.4% / 3 [5] (few samples)<br>94% / 7 vehicles [6], Data fusion with inertial sensors | low/ medium | no | yes | yes | N |
| Camera-vision | 88.11%-95.7% / 6 [7] (multiple datasets) | medium | no | yes | no | DLW |
| Inertial Sensors | 99% / 3 [8], Accelerometer / Magnetometer (few samples)<br>93.4% / 3 [9], Accelerometer / Magnetometer | low / high | possible | yes | yes | T |
| Induction Loop | 76.4%-99% / 3 [10] (class-specific accuracy)<br>99% / 3 [11] (dual-loop setup) | high | yes | yes | yes | J |
| GPS-trajectories | 95.79% / 2 [12] (post-processing analysis of the acceleration) | low | no | no | no | |
| Laser Scanning | 91.8% / 3 [13] (few samples)<br>86% / 2 [14] | high | yes | yes | yes | DW |
| Piezoelectric | 88.33%-97.35%% / 10 [15] (multiple locations, few samples) | high | yes | yes | yes | ST |
| Radar | 85% / 5 [16] | low | no | yes | yes | J |
| WIM | 90% / 13 [17] | very high | yes | yes | yes | J |
| **Radio-fingerprinting** | 99% / 2, 89.15% / 9 (**this paper**) | low | no | yes | yes | |

*Vulnerabilities*: D: Dirt, J: Traffic Jam, L: Lightning, N: Noise, S: Vehicle Speed, T: Temperature, W: Weather

count related 13-category Federal Highway Administration (FHWA) scheme F exist, most research works introduce their own classification schemes in order to achieve a better match for the intended application. Consequently, comparing the achieved accuracy results is a non-trivial task. Since the considered approaches differ in the classification granularity, it should be denoted that the required number of classes is highly depending on the actual intended application. While traffic analysis systems benefit from fine-grained taxonomies, binary classification is sufficient for mostly length-dependent use-cases such as parking space accounting systems.

A general overview of well-established real-world sensor systems, e.g. induction loops and WIM, is provided by the FHWA in [17]. Additionally, in [12], a detailed summary about existing approaches is provided and an exotic approach for offline analysis of the acceleration behavior derived from Global Positioning System (GPS)-traces is proposed. The following paragraphs provide some additional information about the directions of current research work and achieved results.

*1) Acoustic Signals:* Classification based on acoustic signals has recently achieved great attention in other research fields, e.g. the authors of [18] propose an acoustics-based classification mechanism for flying insects. Different approaches have been proposed to apply the existing methods in the vehicular domain. The authors of [5] apply an acoustics-based approach in the frequency domain using Artificial Neural Network (ANN) and k-Nearest-Neighbor (KNN). Extensive preprocessing is required to remove audio noise such as the horn signal. Nevertheless, the achieved Classification Success Ratio (CSR) is 73.4% using ANN. In [6], a sensor fusion approach is proposed using acoustics, magnetometers and electromagnetic radio field sensing. In contrast to most other approaches, the authors do not classify vehicles into abstract classes but aim to identify the actual vehicles themselves by their fingerprints.

*2) Inertial Sensors: Accelerometers and Magnetometers:* The use of inertial sensors, such as accelerometers and magnetometers, is a further approach for classifying vehicles. While the actual sensors are cheap, the resulting cost-efficiency of the corresponding classification systems highly depends on whether the installation is road-intrusive and requires additional roadwork. Often, different inertial sensors are combined in order to exploit their unique advantages. The authors of [8] present a vehicle classification system using an installation combining multiple magnetometers for speed estimation and accelerometers for determination of the number and the spacing of the vehicle's axles. While their proposed system is experimentally evaluated and validated with a commercial WIM system, the number of evaluation samples is relatively low, especially for specific vehicle classes. In [9] a similar approach is presented that is using ANN, Support Vector Machine (SVM) and Logistic Regression (LR) for differentiating three vehicle classes which achieves an accuracy of 93.4% on a large data set with LR.

*3) Vision-based Approaches:* Video-based classification approaches exploit the wide availability of traffic cameras. Since the research field of machine learning on image data has made great advances in the recent years, existing classification methods from adjacent research fields can be exploited and often directly applied. As large reference datasets are available, many research works address fine-grained classification. However, the performance is highly reduced in the presence of vision obstructions caused by dirt or weather conditions and is highly depended on the lightning conditions. Additionally, this approach system-immanently has to deal with privacy-related challenges. A detailed summary of using traffic cameras for detecting and tracking of vehicles is given in [19]. In [14], the authors propose a semi-supervised Convolutional Neural Network (CNN) approach for differentiating six vehicle classes that achieves a CSR of 95.7% on a daylight dataset and a classification accuracy of 88.8% on a nightlight set.

*4) Radio-fingerprinting / Radio Tomography:* The proposed radio-fingerprinting approach leverages the idea of Radio Tomographic Imaging (RTI) [20] to classify cars based on their signal attenuation characteristics while the vehicles pass an installation of communicating Wireless Sensor Network (WSN) nodes. Since the system relies on off-the-shelf hardware and does not involve roadwork, it can cost-efficiently be deployed in an ad-hoc manner. Moreover, it is robust against challenging environmental conditions and is even able to cope with hardware failures due to the system-immanent redundancy by using multiple radio links for the machine learning-based classification.

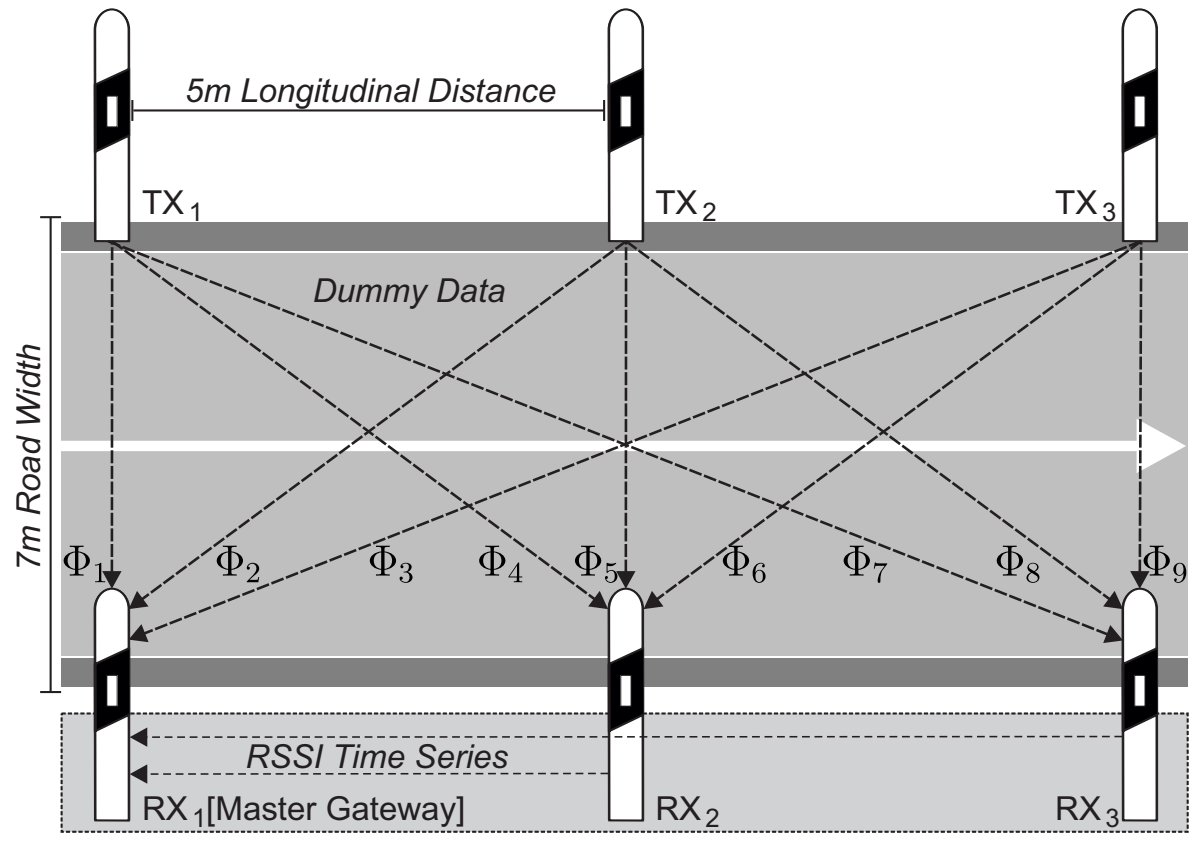

Fig. 2. Schematic overview of the proposed radio-based vehicle classification system illustrating the different radio links used for the vehicle classification.

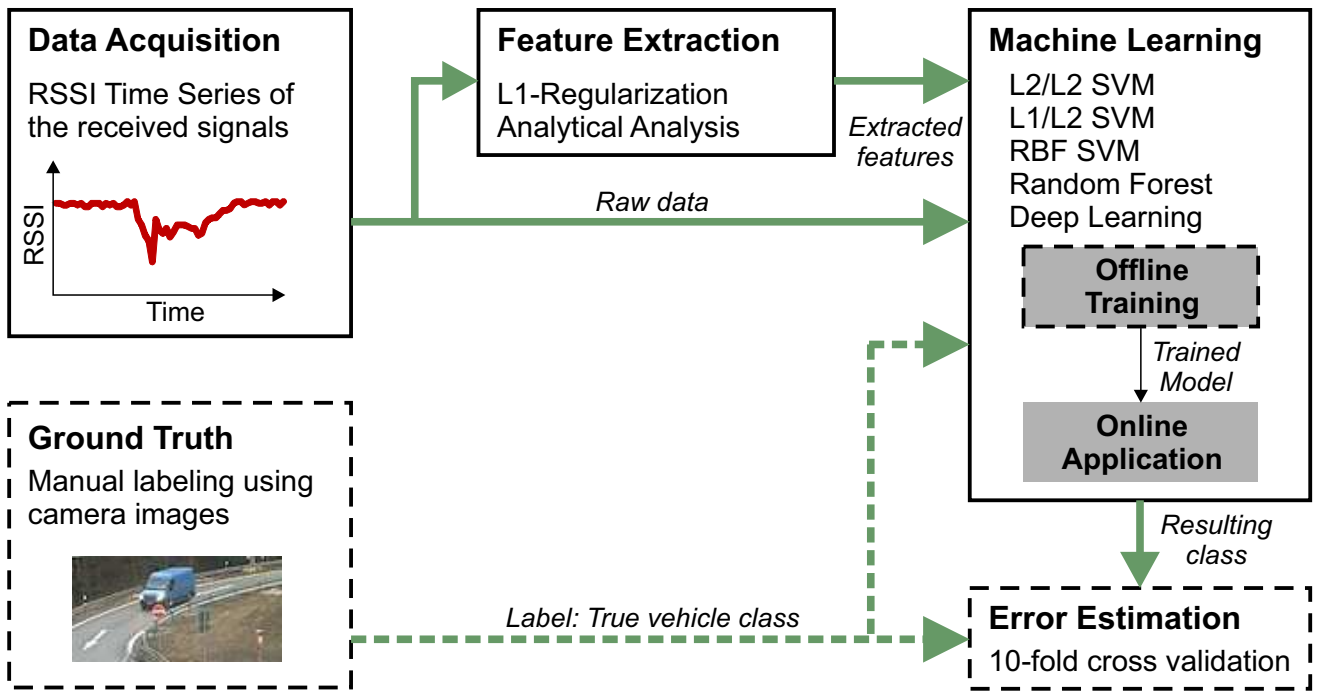

Fig. 4. Processing pipeline for the machine learning-based vehicle classification. The dashed components are only required during the offline training and evaluation phase.

## III. Machine Learning Based Solution Approach

A schematic overview of the key system components is shown in Fig. 2. Three delineator posts equipped with IEEE 802.15.4 low power radio modules are installed on each side of the street within fixed longitudinal distances. One road side contains all transmitter nodes that transmit beacons using a token-based channel access method. The other road side is used for the receivers. Overall, nine different radio links $\Phi_i$ exist in the system, since each of the receivers receives all transmitted signals. All receiver nodes forward their corresponding measurements of the Received Signal Strength Indicator (RSSI) to the master gateway that is responsible for performing the actual classification task and is the gateway for the on-side data exploitation.

Fig. 3 shows an example comparison of the attenuation behavior for a passenger car and a truck. The characteristics in terms of drop duration, attenuation magnitude and overall attenuation duration differ significantly between the two classes. It can be concluded, that different vehicle types have unique radio-fingerprints, that depend onW their shape and material. While the main focus of this paper is the binary vehicle classification, the applicability of the proposed system is also evaluated for a fine-grained classification task with nine classes. Therefore, a hierarchical application-driven classification scheme is applied, which is similar to the Level 2 classification of the Nordic system for intelligent classification of vehicles (NorSIKT) applied in [9] for the coarse-grained binary classification. Tab. II summarizes the resulting taxonomy and provides a mapping to the FHWA classes for the coarse-grained binary classification.

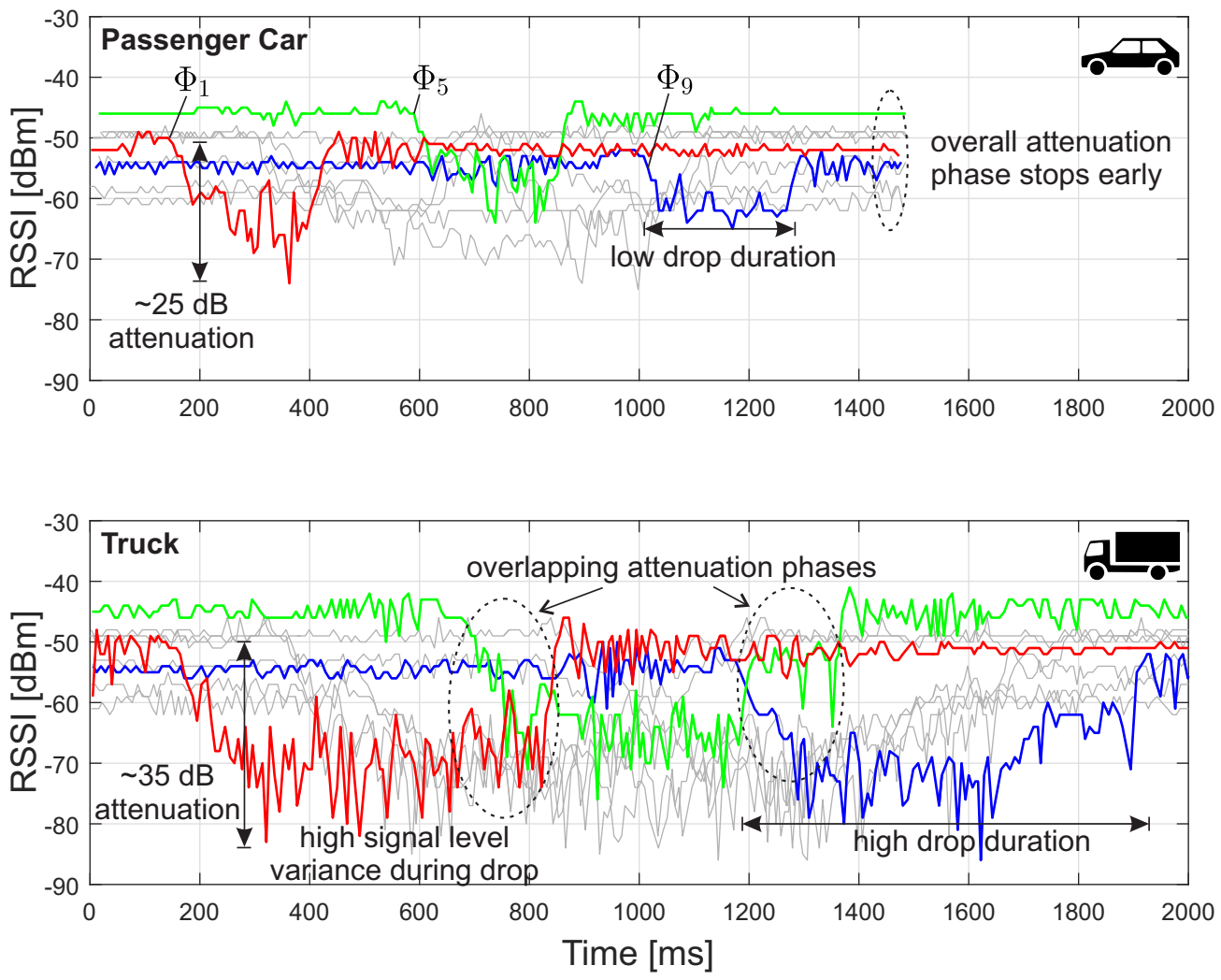

Fig. 3. Example comparison of multidimensional radio-fingerprinting for cars and trucks in the time domain. Highlighted are the signals for the three direct radio links $\Phi_1$, $\Phi_5$ and $\Phi_9$.

TABLE II
TAXONOMY

| Coarse-grained Class | Fine-grained Class (Number of Samples) |
|---|---|
| **Car-like** *FHWA Classes* 1-3,5 | Passenger car (1528)<br>Passenger car with trailer (19)<br>SUV (93)<br>Minivan (128)<br>Van (172) |
| **Truck-like** *FHWA Classes* 4,6-13 | Truck (75)<br>Truck with trailer (52)<br>Bus (5)<br>Semi truck (563) |

The processing pipeline for the machine learning-based classification is illustrated in Fig. 4. After the data acquisition phase, where the time series of the RSSI of all radio links is collected, different feature extraction methods are applied in order to identify features that describe the relevant signal characteristics in a more memory-efficient way than the raw signal data. Afterwards, different machine learning models are trained and their classification results are compared to the true vehicle classes in a 10-fold cross validation.

## IV. Methodology

In this section, the setup for the experimental evaluation as well as the considered performance indicators and machine learning methods are presented.

Tab. III denotes the system design parameters. The raw data of the classification results for all methods is publicly available at [21]. Each data point consists of nine time series $\Phi_1, \Phi_2, \ldots, \Phi_9$ that each contain 800 RSSI samples.

TABLE III

SYSTEM DESIGN PARAMETERS

| Parameter | Value |
|---|---|
| Transmission power | 2.5 dBm |
| Operating frequency | 2.4 GHz |
| Antenna type | Omnidirectional |
| Antenna height | 1 m |
| Road width | 7 m |
| Longitudinal post distance | 5 m |

### A. *Field Deployment Setup*

For the following evaluations, we analyzed measurement data from 2635 vehicles which was obtained on an experimental live deployment of the proposed system at the entrance of a rest area on the German highway A9 within an official test field by the German Federal Ministry of Transport and Digital Infrastructure (c.f. Fig. 5). The ground truth labels for the machine learning process were manually determined based on video data.

### B. *Quality Measure and Machine Learning Models*

The quality of a machine learning model $f$ on a dataset $\mathcal{D}$ via the correct-classification-ratio a.k.a. *accuracy* is measured as:

$$\mathrm{ACC}(f;\mathcal{D}) = \frac{1}{|\mathcal{D}|} \sum_{(y,\boldsymbol{x})\in\mathcal{D}} 1_{\{y=f(\boldsymbol{x})\}} \ .$$

Here, $|\mathcal{D}|$ is the cardinality of $\mathcal{D}$, and $1_{\{y=f(\boldsymbol{x})\}}$ is the indicator function that only evaluates to 1 if $f(\boldsymbol{x})$ outputs the correct class $y$, and 0 otherwise. When the cardinality of $\mathcal{D}$ approaches $\infty$, $\mathrm{ACC}(f;\mathcal{D})$ converges to the probability of doing a correct classification, i.e., $\lim_{|\mathcal{D}|\to\infty} \mathrm{ACC}(f;\mathcal{D}) = \mathbb{P}(f(\boldsymbol{x}) = Y)$. However, the value of $\mathrm{ACC}(f;\mathcal{D})$ will only be a reliable estimate to $\mathbb{P}(f(\boldsymbol{x}) = Y)$ when $\mathcal{D}$ is an "unseen" data set that was not used to learn $f$. Our data is thus partitioned into a training set $\mathcal{D}_{\text{train}}$, which is used to learn the model $f$, and a test set $\mathcal{D}_{\text{test}}$, which is used to assess the quality of $f$. Nevertheless, we might be unlucky with a particular choice of $\mathcal{D}_{\text{train}}$ and $\mathcal{D}_{\text{test}}$, resulting in overly confident or overly pessimistic estimates $\mathbb{P}(f(\boldsymbol{x}) = Y)$. To overcome this issue, the data set $\mathcal{D}$ is partitioned into $k$ sets $\mathcal{D}_1, \dots, \mathcal{D}_k$. The learning procedure is then repeated $k$ times, where in each run, the set $\mathcal{D}_i$ takes the role of the test set to compute $\mathrm{ACC}(f;\mathcal{D}_i)$ and all remaining data sets are used for training. The $k$ accuracy values are averaged and the resulting quantitiy is the $k$-fold cross-validated accuracy $\mathrm{ACC}_k = (1/k)\sum_{i=1}^{k} \mathrm{ACC}(f;\mathcal{D}_i)$ [22]. On the one hand, the cross-validated accuracy is more reliable, since it reduces the chance of choosing a train/test split which results in an artificially high or low accuracy. On the other hand, it allows to estimate the standard deviation:

$$\hat{\sigma}_{\mathrm{ACC}} = \sqrt{\left(\frac{1}{k}\sum_{i=1}^{k} \mathrm{ACC}(f;\mathcal{D}_i)^2\right) - \mathrm{ACC}_k^2}$$

of the $k$-folds, which in turn quantifies the uncertainty of the estimate. In this paper, $k = 10$ is used.

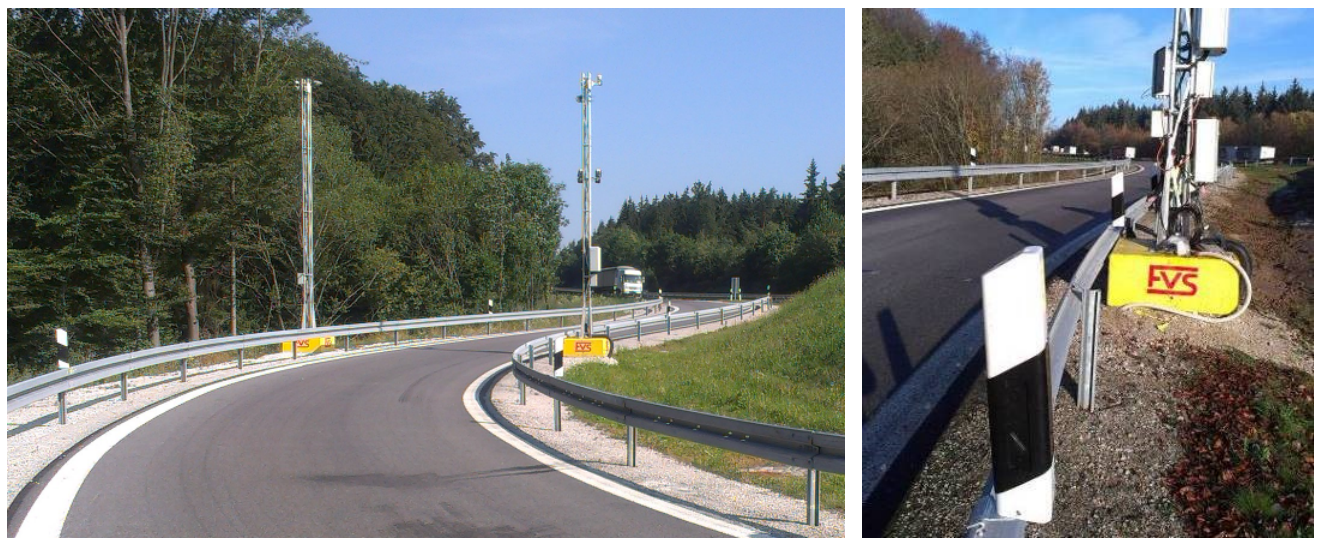

Fig. 5. Experimental live deployment of the system on the German highway A9 within an official test field by the German Federal Ministry of Transport and Digital Infrastructure.

### C. *Classification Methods*

Various machine learning models are qualified for the classification task. The most important model classes are investigated in the following: Support Vector Machines (SVM) [23], Deep Learning (DL) via Convolutional Neural Networks [24], and Random Forests (RF) [25].

*a) SVM:* Support vector machines are designed to separate data points in a $d$-dimensional real-valued space by estimating a hyperplane that cuts the $\mathbb{R}^d$ into half spaces, such that the majority of points which belong to the first class are on one side of the hyperplane while the majority of the other points are on the other side. To apply SVMs to more than two classes, the one-vs-all strategy is employed. Details about the implementation can be found in [26]. Classification is achieved via $f(\boldsymbol{x}) = \operatorname{sign}(\langle\boldsymbol{\beta},\boldsymbol{x}\rangle)$ with weight vector $\boldsymbol{\beta} \in \mathbb{R}^d$ (to simplify the notation, the bias term is fixed to $\boldsymbol{\beta}_0 = 0$), and the hyperplane is learned by minimizing a specific objective function. Here, $d = 9 \times 800 = 7200$ for the raw data, i.e., the raw values of all nine time series are sequentially stacked into one vector. However, different variants of the objective allow to learn SVMs with different properties: The *$\ell_2$-regularized $\ell_2$-loss support vector machine* (L2/L2 SVM) is the classic linear SVM with the so-called squared hinge loss. Its objective function can be written as

$$\min_{\boldsymbol{\beta}} \frac{1}{2}\|\boldsymbol{\beta}\|_2^2 + C \sum_{(y,\boldsymbol{x})\in\mathcal{D}} \underbrace{\max\{0, 1 - y\langle\boldsymbol{\beta},\boldsymbol{x}\rangle\}^2}_{\text{Squared Hinge-Loss}} \ .$$

The parameter $C$ allows to trade the importance of correctly classified training data points against the norm of the weight vector (model complexity). Note that it is not necessarily desirable to achieve a very low error on the training data, since this can result in *overfitting*, i.e., a low generalization performance on unseen data. By altering the first term from $\ell_2$ to the $\ell_1$-norm, the L1/L2 SVM objective is recovered:

$$\min_{\boldsymbol{\beta}} \|\boldsymbol{\beta}\|_1 + C \sum_{(y,\boldsymbol{x})\in\mathcal{D}} \max\{0, 1 - y\langle\boldsymbol{\beta},\boldsymbol{x}\rangle\}^2 \ .$$

While the $\ell_2$-norm prevents the model weights from becoming large, the $\ell_1$-norm prefers sparse models, i.e., weight vectors which contain many 0 values. This property is particularly important if the relevance of specific data dimensions (*features*) shall be assessed for the classification task, irrelevant features will exhibit zero-valued weights. In contrast to the first two models, the third type of SVM model that is investigated, employs a non-linear feature transformation to perform the classification of non-linearly separable data in a high-dimensional feature space. The explicit computation of high-dimensional features is circumvented by the so-called "kernel trick". Here, the radial basis function (RBF) kernel $K(\boldsymbol{x}^i, \boldsymbol{x}^j) = \exp(-\gamma\|\boldsymbol{x}^i - \boldsymbol{x}^j\|_2^2)$ is employed.

*b) RF:* Random forests [27] are sets of bootstrapped [28] decision trees. Suppose that $T = (V, E)$ is a directed, tree structured graph. Each vertex $v \in V$ has (at most) two children $v_{\text{left}}$ and $v_{\text{right}}$. Moreover, for each vertex $v$, $\text{val}(v) \in \mathbb{R}$ is a real number, $\text{idx}(v) \in \{1, 2, \ldots, d\}$ is a feature index, and $\text{ch}(v)$ returns the number of children. The vertex function $f(v, \boldsymbol{x})$ is then

$$f(v,\boldsymbol{x}) = \begin{cases} f(v_{\text{left}}, \boldsymbol{x}) & , \boldsymbol{x}_{\text{idx}(v)} \leq \text{val}(v) \wedge \text{ch}(v) = 2 \\ f(v_{\text{right}}, \boldsymbol{x}) & , \boldsymbol{x}_{\text{idx}(v)} > \text{val}(v) \wedge \text{ch}(v) = 2 \\ \text{val}(v) & , \text{else} \ . \end{cases}$$

The prediction function of a single decision tree $T$ with root $v_0$ can then be written as $T(\boldsymbol{x}) = f(v_0, \boldsymbol{x})$. Solving the corresponding learning problem is likely to have exponential runtime complexity. Thus, practical decision trees are grown heuristically by sequentially choosing $\text{idx}(v)$ and $\text{val}(v)$ such that the mutual information is maximized. Finally, a random forest is a set of decision trees $\mathcal{T} = \{T_1, T_2, \ldots\}$, where each tree $T_i$ is grown on a random bootstrap sample $\mathcal{D}_i$ of the training data and a random feature set. The prediction of the forest is then determined via majority voting, i.e., the forest predicts the class which is predicted by most of its trees.

*c) Deep Learning:* Deep convolutional neural networks [29] are the machine learning technique which is currently receiving great attention from the research community and beyond. While classic (non-deep) approaches rely on hand-crafted features, various hyper parameters, or user defined feature transformations (e.g., the RBF kernel), deep learning methods aim at phrasing almost all parts of the model as differentiable function. Thus, numerical optimization methods can replace what was formerly done manually. These methods work especially well in computer vision tasks, where a large number of semantically equivalent features is present, e.g., the pixel colors of an image. Convolutional networks exploit the spatial locality of pixel features by moving small *filter matrices* over the image, performing a convolution operation with the underlying pixel values. By using $k$ filter matrices, the convolutional network learn to detect $k$ different patterns in the input image, independent of the position of the pattern. Since the output of such a convolutional layer is again tensor shaped, a similar convolution may be stacked on top, forming a (deep) network of convolutional layers. The final output is fed into a classic multi layer perceptron.

In the applied setting, the raw features consist of 9 time-series - one per radio link - each of length 800. Arranging the features in a $9 \times 800$ matrix reveals a similarity to image data. While $9 \times 800$ is no meaningful resolution for real-world images, deep learning methods process any matrix or tensor shaped object. Hence, filters may be moved over the time series data to detect patterns which in turn serve as features for the classification.

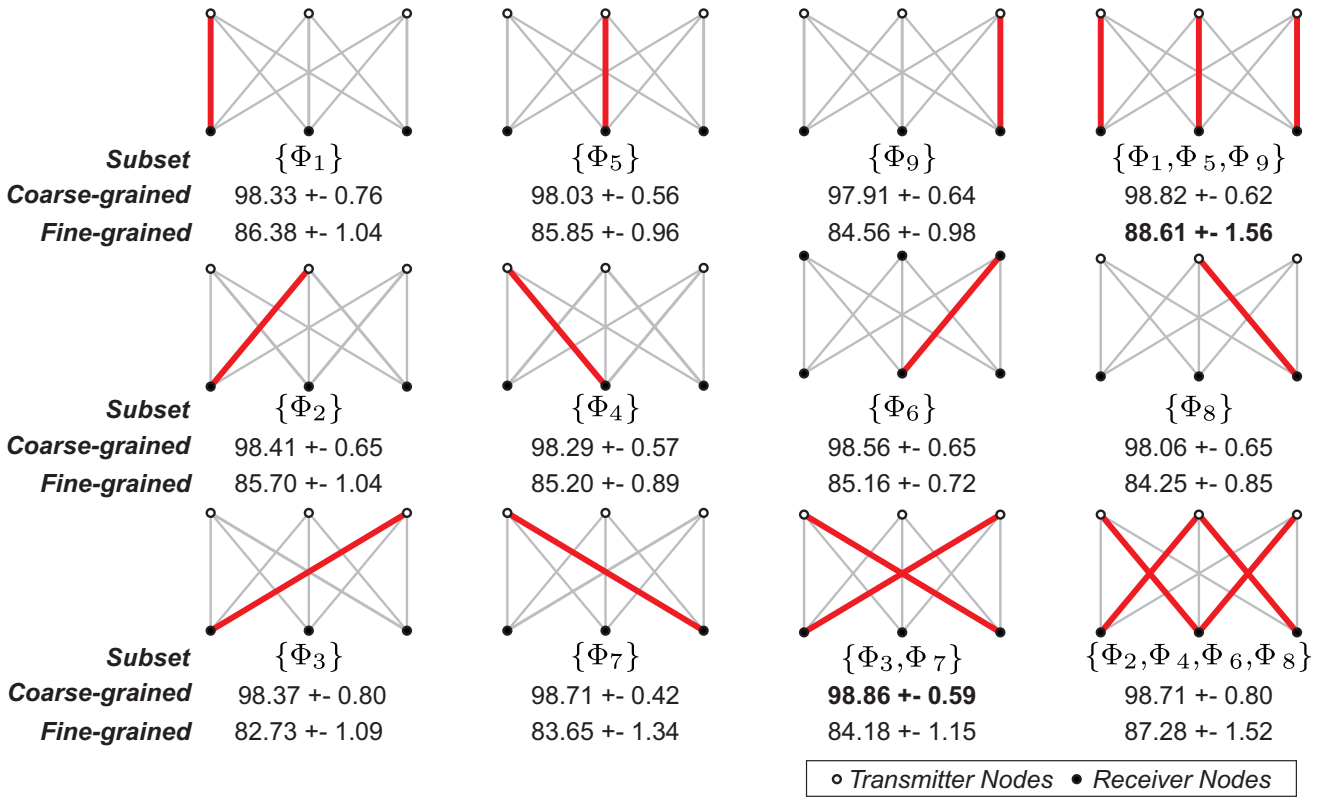

Fig. 6. Overall classification accuracy with RBF SVM for different subsets of radio links. The active links are highlighted in red.

## V. Results

In this section, the results obtained from the experimental evaluation campaign are presented and discussed. Furthermore, the resource-efficiency of the considered machine learning approaches is evaluated from an embedded devices perspective.

The classification models for L2/L2 and L1/L2 SVMs are learned with `liblinear`[1], the SVM with RBF kernel is learned with `libsvm`[2], RF models are produced with `rapidminer`[3], and the deep learning models are trained with an implementation that is based on `tiny-dnn`[4].

We conducted a grid search to find the hyper-parameter values: For the linear SVMs, the default value $C = 1$ provided reasonable results, for the RBF kernel SVM, $C = 10$ and $\gamma = 10^{-2}$ provided the best results. The random forest is generated with 128 trees, each of maximum depth 32 as more or deeper trees did not provide better results. While deep learning methods enjoy automatic tuning of structured hyper parameters, like feature-tranformations, the user has to specify the network architecture. Multiple designs were evaluated, where the best results are delivered by one convolutional layer with 16 filters of size $9 \times 20$, that is, each filter moves over 20 consecutive time steps of all nine time series simultaneously. The output is fed into a rectified linear activation unit (ReLU), followed by an average pooling layer and three fully connected ReLU layers. The final output is generated by softmax units which allows the interpretation of the network outputs as pseudo probabilities for each class. Batch normalization is applied after the first fully connected layer. The model weights are trained with the *ADAM* algorithm [30] via multiclass cross-entropy minimization. Details on deep learning architectures can be found in [29].

### A. Feature Importance

Determining the relevance of the considered features is the first step to get an intuition about the expectable classification performance. Moreover, analyzing the importance of radio links $\Phi_i$ can improve the efficiency of the measurement

[1] `https://www.csie.ntu.edu.tw/~cjlin/liblinear`
[2] `https://www.csie.ntu.edu.tw/~cjlin/libsvm`
[3] `https://rapidminer.com`
[4] `https://github.com/tiny-dnn/tiny-dnn`

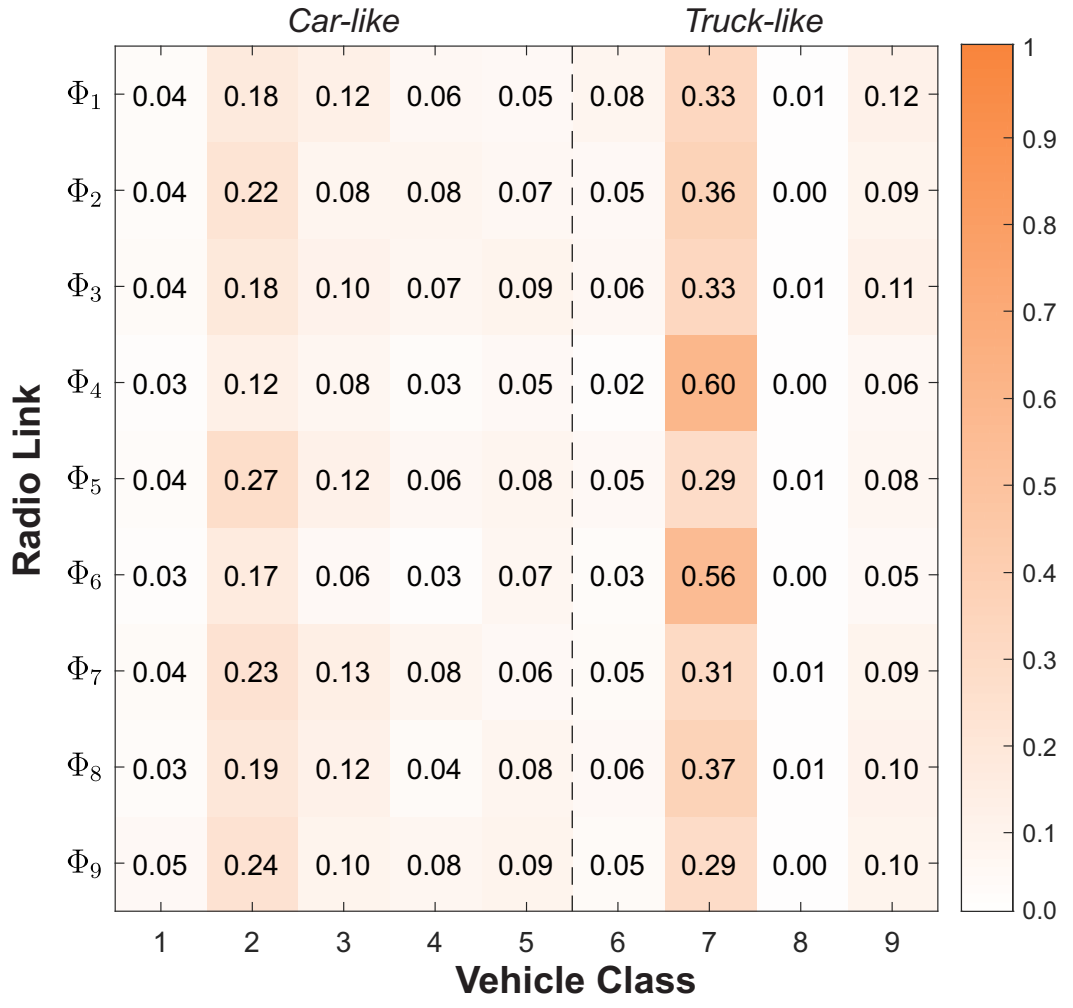

Fig. 7. Relative importance of the individual radio links per vehicle class derived from the nuber of non-zero features per radio link per vehicle class.

setup, by identifying links which can be dropped without loosing a significant amount of accuracy.

First of all, the accuracy that can be achieved with single links and several canonical subsets of links is evaluated. The corresponding results are shown in Fig. 6. Due to space constraints, only the results of the RBF SVM are presented, which achieves the highest overall accuracy. In case of the fine-grained (nine classes) classification problem, it can be seen that the three direct links $\Phi_1, \Phi_5, \Phi_9$ suffice to deliver a 10-fold cross validated accuracy of $> 88\%$—the estimated standard deviation amounts to $1.56\%$. For the coarse-grained scenario (binary), the two long-range diagonal links provide the best performance. Nevertheless, all settings achieved more than $98\%$ cross-validated accuracy. Considering the estimated standard deviations, it can be concluded that the system is highly redundant and could be deployed with a reduced setup in order to increase the cost-efficiency. For example, subset $\{3, 7\}$ achieves the highest accuracy for binary classification with only four involved delineator posts. To get a more detailed view on the importance of each link for each class, the inherent feature selection of L1/L2 SVMs is exploited. Recall that irrelevant weights are pinned to $0$ by this model type. A classification model is learned on the whole data set. Due to one-vs-all multiclass classification, this results in one weight vector $\boldsymbol{\beta}_y \in \mathbb{R}^{7200}$ for each class $y$ with 800 weights per link. For each combination of link and class, the number of non-zero weights is counted. The resulting quantities, normalized for each link, are shown in Fig. 7. The figure reveals the relative importance of each link for the fine-grained classification. Colors with higher intensity indicate that many time steps of each time series are required for the classifier. Classes 1-9 obey the same order as the fine-grained classes from the taxonomy (Tab. II). Moreover, classes which require only a relatively low number of weights of each link (e.g., classes 1, 6, and 8) are *easier* to learn than those which consistently require a high number of weights (e.g., classes 2 and 7). Especially $\Phi_4$ and $\Phi_6$ are relevant for the classification of trucks with trailers (class 7). For now, the relevance of links based on raw feature values was investigated. To understand in which way the shape of time series affects the classification results, the dynamic time warping (DTW) distance between the radio link time series for each pair of training data points is considered. DTW is a specialized distance measure for time series which is designed to detect if time series are similar, even if their peaks are slightly shifted or squeezed [31]. For easy interpretation, the DTW distance is converted to a similarity measure via $\exp(-\overline{\mathrm{dtw}}(i, j))$, where $\overline{\mathrm{dtw}}(i, j)$ is the standardized DTW distance between the $i$-th and $j$-th data point. The resulting similarity matrices are shown in Fig. 8. Bright color indicate similarity while darker colors indicate dissimilarity. Each diagonal block corresponds to a fine-grained class and the red lines indicate the diagonal blocks that correspond to the two coarse-grained classes. It can be seen that all matrices contain bright areas whithin some diagonal blocks, i.e., the DTW distance seems to be suitable for classification. However, when these matrices were applied as kernels in an SVM model, e.g., via multiple kernel learning [32], only $57.989\%$ accuracy could be achieved. We conclude that analytic features of time series which are ignored by the DTW distance must be important for the classification. The DTW distance between time series might be large, while other characteristics which are relevant for the classification are similar.

We identified simple analytic features for each time series: the positions and values of the three largest and smallest measurements in each series (top-3 minima and maxima) as well as the area, sum of squares, and sum of cubes, e.g., $A(i, j) = \sum_{t=1}^{800} \Phi_i(t)^j$ for $j \in \{1, 2, 3\}$. Those will be referred to as the *reduced* features, while the original time series data contains the *raw* features. In the following, both feature sets, *raw* and *reduced*, are considered.

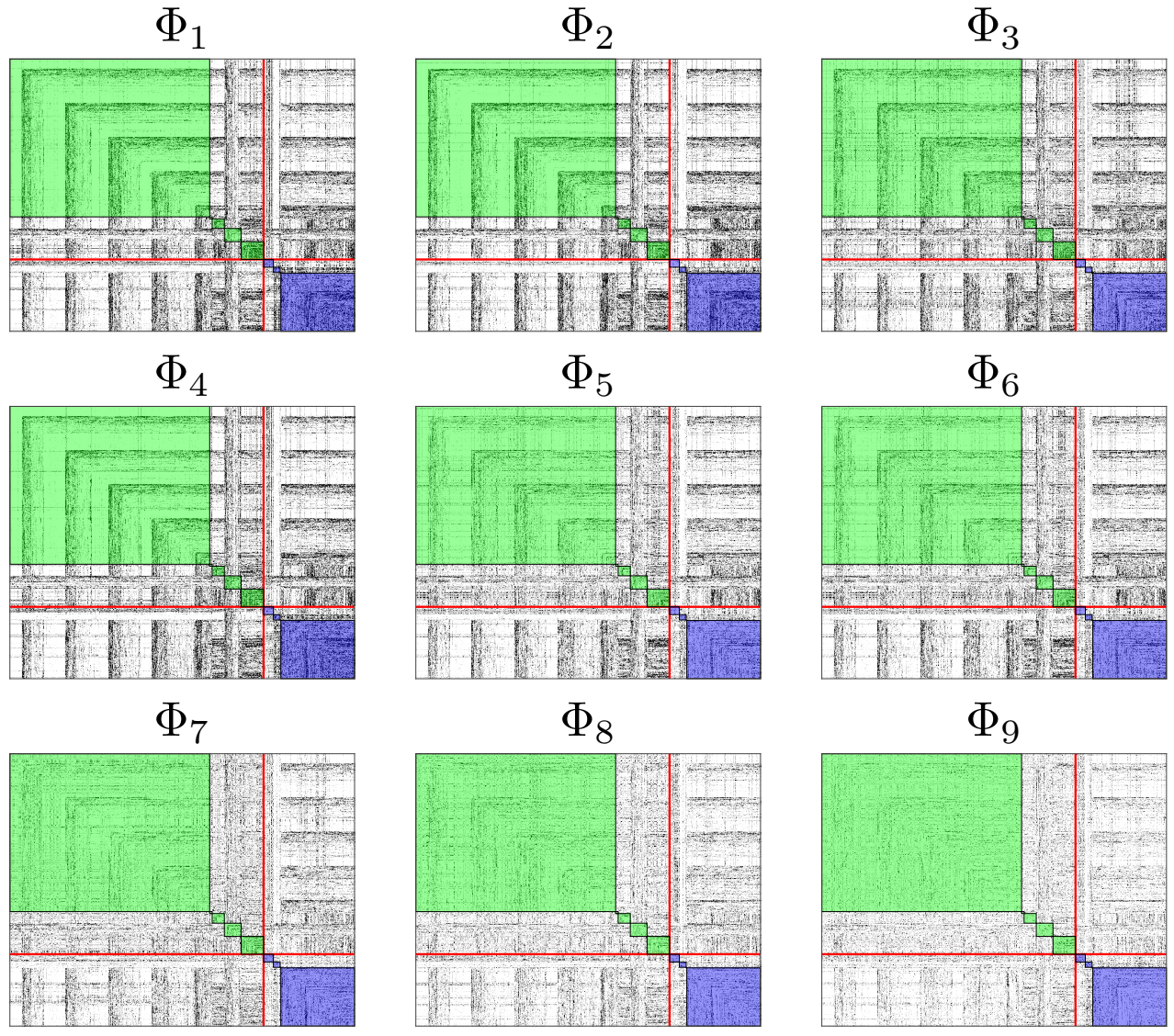

Fig. 8. Exponentiated negative dynamic time warping similarity between each pair of data points for each radio link.

## B. Classification Performance

In the following paragraph, the classification performance is evaluated. Cross-validated accuracy and the corresponding standard deviation for the coarse- and fine-grained classification problems are shown in Fig. 9. While all methods achieve rather similar results, the only one that sticks out is the RBF SVM which achieves over 99% accuracy on average for the coarse two-class problem. Regarding all nine classes, 89% are achieved on average. However, only 1392 weights are required for the L1/L2 SVM which achieves more than 98.8% accuracy on the two-class problem. This is remarkable, since the RBF SVM has more than $10^6$ parameters since kernel SVMs store full data points to construct the hyperplane. The RBF SVM model consists of 179 *support vectors*, each of 7200 dimensions which results in more than $10^6$ weights. However, the most resource-efficient result is delivered by the L1/L2 SVM on the *reduced* features: only 98 non-zero weights are required to achieve 98.2% accuracy. Even ultra-low-power systems offer enough resources to run this model in real-time.

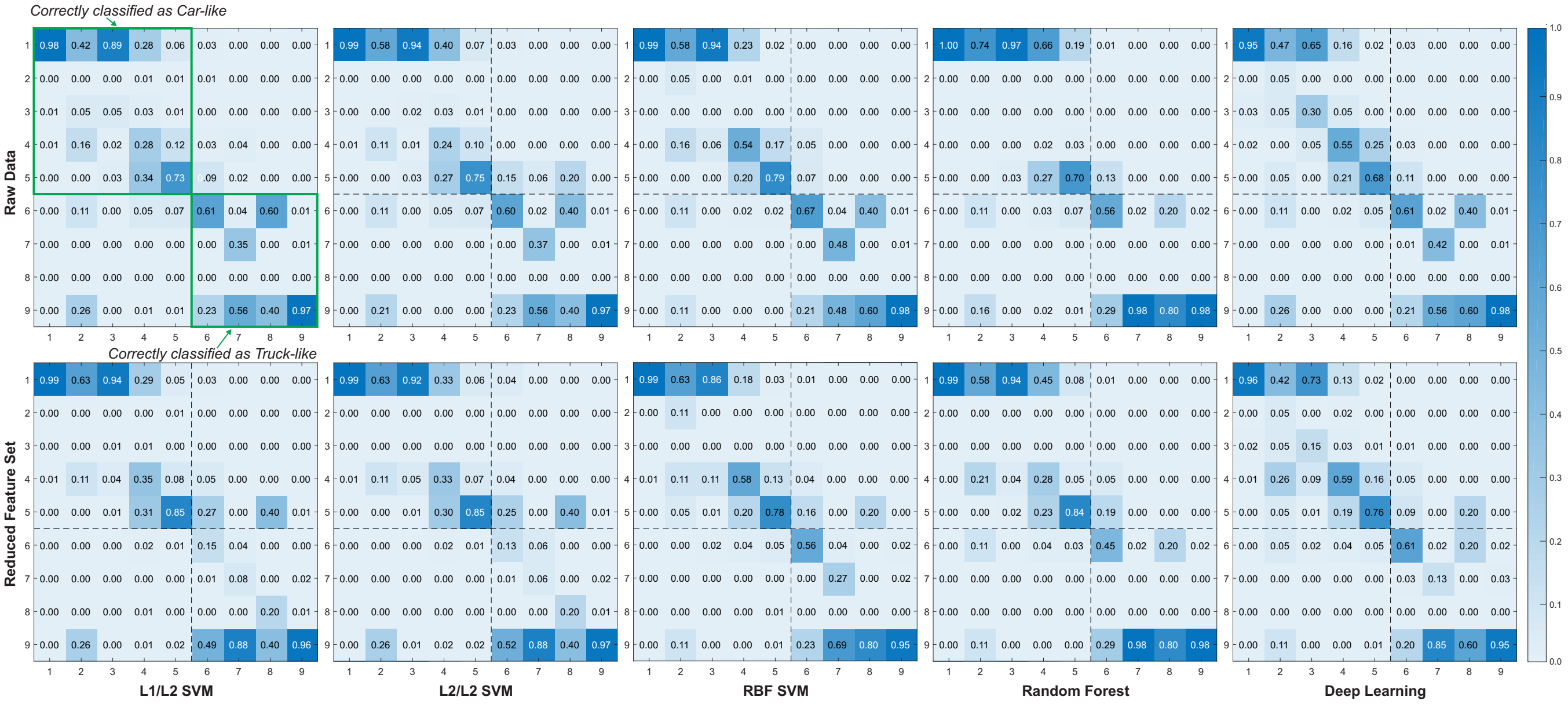

Fig. 10. Normalized confusion matrices for the fine-grained classification task. *Vehicle classes: 1: Passenger car, 2: Passenger car with trailer, 3: SUV, 4: Minivan, 5: Van, 6: Truck, 7: Truck with trailer, 8: Bus, 9: Transporter.*

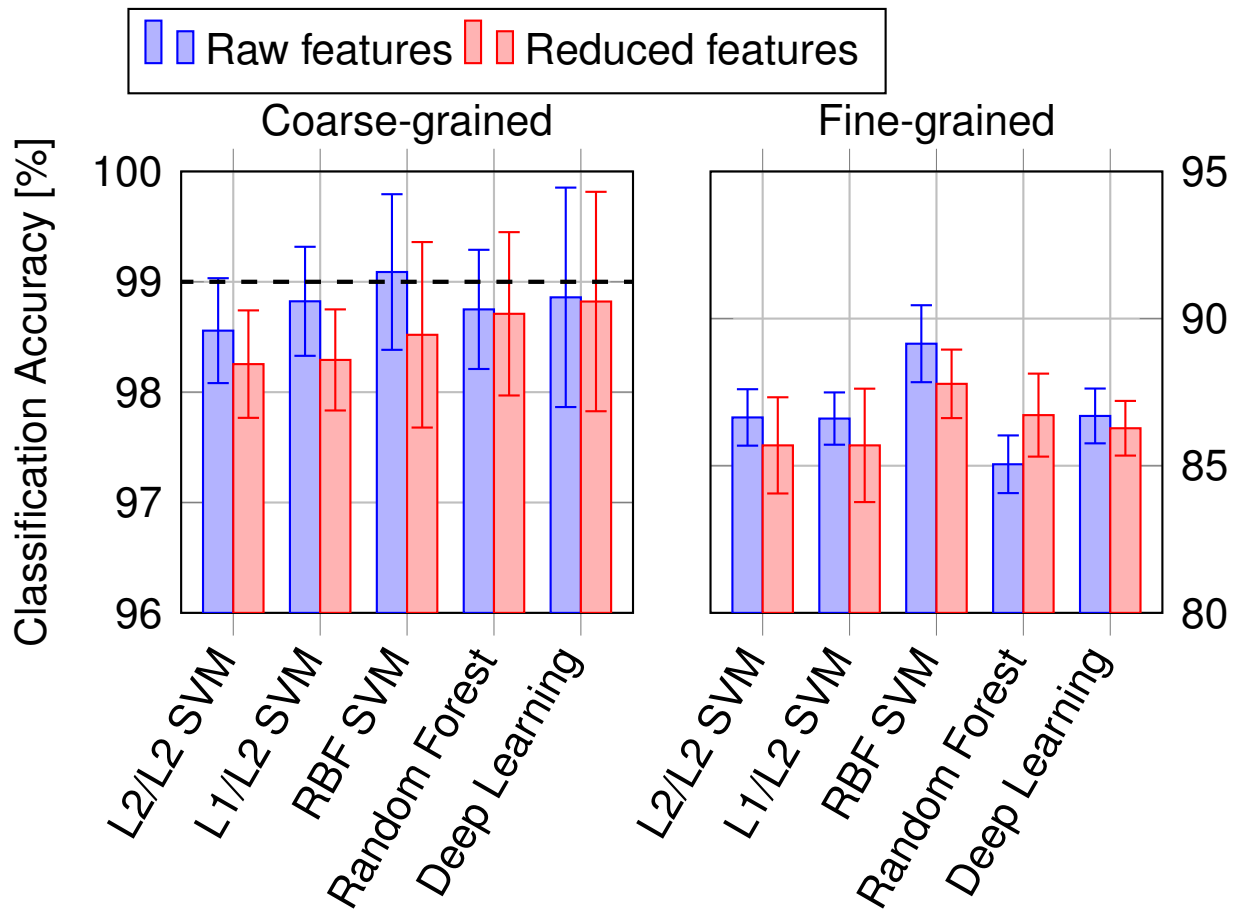

Fig. 9. Cross-validated accuracy of each machine learning model (L1/L2 SVM, L2/L2 SVM, RBF SVM, Random Forest, and Deep Learning), trained on *raw* and *reduced* feature sets.

To get a more detailed understanding of the quality of the models, the class-wise performance via confusion matrices shown in Fig. 10 and Fig. 11 is investigated. While all methods show a similar general behavior, the highest accuracy is again achieved with the RBF SVM. For the Sport Utility Vehicle (SUV) and the bus subclass, the fine-grained classification fails due to the high similarity to their respective most significant elements within the coarse-grained class and the relatively low sample size of SUVs and buses. Nevertheless, most fine-grained miss-classifications do not lead to coarse-grained miss-classifications. As Tab. IV shows, the coarse-grained miss-classifications for the car-like class are mainly caused by cars with trailers that have an overall length which is rather typical for elements of the truck-like class.

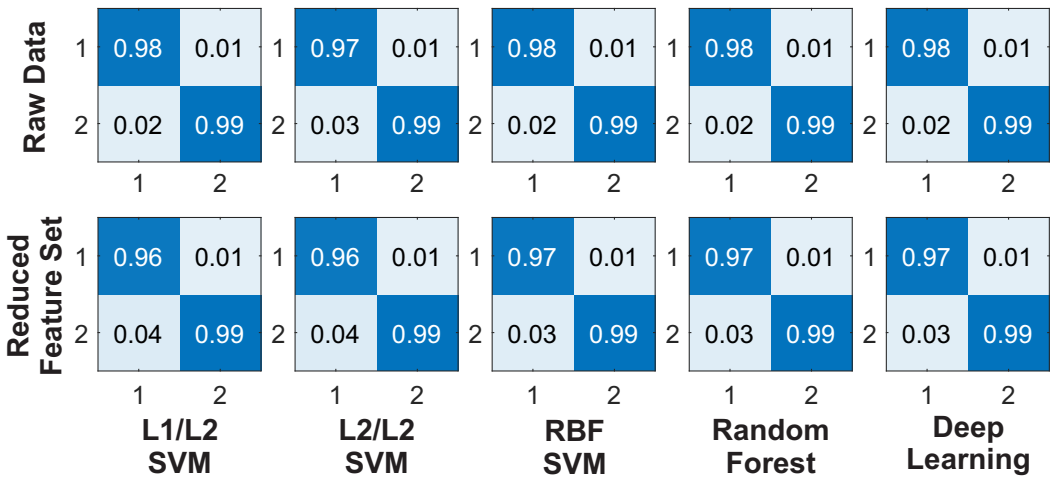

Fig. 11. Normalized confusion matrices for binary classification task. *Vehicle classes: 1: Car-like, 2 Truck-like.*

## VI. Conclusion

Up-to-date information about vehicle types is highly important for next generation ITSs in a smart city context and enables services such as smart parking and type-specific intelligent traffic control. In this paper, we presented a novel vehicle classification system based on multidimensional radio-fingerprinting which system-immanently provides a favorable set of system properties such as high cost-efficiency, privacy-preservation and robustness in challenging environments. In comprehensive field evaluations within an experimental live deployment, the proposed system achieved more than 99% accuracy for binary classification of cars and trucks and 89.15% accuracy for a fine-grained classification scheme with nine classes using RBF SVM. Different system variants can be configured in order to increase the resource-efficiency (98.82% binary accuracy with L1/L2 SVM) or the cost-efficiency (98.86% binary accuracy with RBF SVM using only two radio links) for the deployment while only having minor impacts on the overall accuracy.
In future work, we will further improve the accuracy by integrating additional sensors into the overall system. In particular, inertial sensors, such as accelerometers and magnetometers, are expected to match well with the system approach and offer the potential of bringing further improvements. Furthermore, more samples for special subclasses will be obtained in order to improve the fine-grained classification accuracy.

TABLE IV
ACCURACY QUOTIENT: MAPPING TO THE CORRESPONDING COARSE-GRAINED CLASS PER FINE-GRAINED CLASS

| Model | Passenger car | Passenger car with trailer | SUV | Minivan | Van | Truck | Truck with trailer | Bus | Semi truck |
|---|---|---|---|---|---|---|---|---|---|
| L1/L2 SVM | **1.00** | 0.63 | **1.00** | 0.94 | 0.92 | 0.84 | 0.94 | **1.00** | **1.00** |
| L2/L2 SVM | **1.00** | 0.68 | **1.00** | 0.95 | 0.93 | 0.83 | 0.94 | 0.80 | **1.00** |
| RBF SVM | **1.00** | **0.79** | **1.00** | **0.98** | **0.98** | **0.88** | **1.00** | **1.00** | **1.00** |
| Random Forest | **1.00** | 0.74 | **1.00** | 0.95 | 0.92 | 0.85 | **1.00** | **1.00** | **1.00** |
| Deep Learning | **1.00** | 0.63 | **1.00** | **0.98** | 0.95 | 0.84 | **1.00** | **1.00** | **1.00** |

## ACKNOWLEDGMENT

Part of the work on this paper has been supported by the German Federal Ministry for Economic Affairs and Energy as part of the cooperation project between Wilhelm Schröder GmbH, TU Dortmund and FH Dortmund (grant agreement number ZF4038101DB5), and by Deutsche Forschungsgemeinschaft (DFG) within the Collaborative Research Center SFB 876 "Providing Information by Resource-Constrained Analysis", projects A1 and B4.